\documentclass[sn-mathphys,pdflatex]{sn-jnl}

\jyear{2021}%

\theoremstyle{thmstyleone}%
\theoremstyle{thmstyletwo}%

\theoremstyle{thmstylethree}%
\newtheorem{definition}{Definition}%

\begin{document}

\title[Gravitational Waves of Type III Shapovalov Spacetimes]{Gravitational Waves of Type III Shapovalov Spacetimes: Particle Trajectories, Geodesic Deviation and Tidal Accelerations}



\author*[1,2]{\fnm{Konstantin} \sur{Osetrin}}\email{osetrin@tspu.edu.ru}

\author[1]{\fnm{Evgeny} \sur{Osetrin}}\email{evgeny.osetrin@tspu.edu.ru}
\equalcont{These authors contributed equally to this work.}

\author[1]{\fnm{Elena} \sur{Osetrina}}\email{elena.osetrina@tspu.edu.ru}
\equalcont{These authors contributed equally to this work.}

\affil*[1]{\orgdiv{Center for Mathematical and Computer Physics}, \orgname{Tomsk State Pedagogical University}, \orgaddress{\street{Kievskaya str. 60}, \city{Tomsk}, \postcode{634061}, 
\country{Russia}}}

\affil[2]{
\orgname{National Research Tomsk State University}, \orgaddress{\street{Lenina pr. 36}, \city{Tomsk}, \postcode{634050}, 
\country{Russia}}}



\abstract{
For gravitational-wave spacetimes of Shapovalov type III, exact general solutions of geodesic deviation equations and equations of motion of test particles are obtained. Solutions are found in a privileged coordinate system, where the metric of the considered spacetime models depends on the wave variable. The exact form of tidal accelerations of the gravitational wave is obtained. In the considered wave models of spacetime, the complete integral of the Hamilton-Jacobi equations of test particles can be constructed. An explicit form of the equations for the transition to a synchronous coordinate system is found, where the proper time of a test particle on the base geodesic is chosen as the time variable, and the time and space variables are separated. In the synchronous coordinate system, the form of the metric of the considered wave spacetime is presented, the form of the geodesic deviation vector and the tidal acceleration vector are obtained. The methods used in the paper and the results obtained are applicable to gravitational waves both in the general theory of relativity and in modified theories of gravity. The proposed approaches are applied to the case of Einstein's vacuum equations.
}




\maketitle

\section{Introduction}
\label{sec1}

%

New approaches proposed recently and the results obtained from the theoretical study of primordial gravitational waves in the Bianchi universes
\cite{Elbistan2021052,Zhang2022035008,Osetrin2022EPJP856,Osetrin2022894,https://doi.org/10.48550/arxiv.2209.08589},
involving obtaining exact models of primordial gravitational waves, obtaining exact solutions to the equations of motion of test particles, exact solutions to the geodesic deviation equations, and exact calculation of tidal accelerations use various mathematical methods, including symmetry theory and the Hamilton-Jacobi formalism. Note that plane gravitational waves in Bianchi universes were started by Vladimir Lukash (see \cite{Elbistan2021052} and references cited in this work).

In this paper, we will consider more general exact models of plane gravitational waves, which include, as a special case, models with spatial homogeneity symmetry that we considered earlier 
\cite{Osetrin2022EPJP856,Osetrin2022894,https://doi.org/10.48550/arxiv.2209.08589}, 
but applicable to a wider class of gravitational waves and allowing exact calculation of the effects created by waves, taking into account the deviation of geodesic and tidal accelerations, both in general relativity and in modified theories of gravity
\cite{Capozziello2021100867,Odintsov2022136817,Odintsov2022100950,Odintsov2022729}, which makes it possible to analytically compare the models and offer observational checks of the realism of the resulting models.

The considered exact models of strong gravitational waves make it possible to calculate both the direct influence of a wave on test particles in order to detect their motion, and to calculate secondary physical effects for waves with large wavelengths, where direct detection is difficult, including the calculation of the radiation of plasma charges in a gravitational wave, the effects of gravitational lensing , calculations of the influence of a passing wave on the change in the periods of pulsars, the capture of physical objects by a gravitational wave, the effect on the observed microwave background, the effect on the observed stochastic gravitational wave background, etc.

The data of direct detection of gravitational waves are currently analyzed using approximate and numerical methods, followed by the formation of computer databases of numerically obtained approximate models. The exact models of gravitational waves obtained can also serve in this case for testing and debugging approximate and numerical methods in this area.

Shapovalov spacetimes \cite{Osetrin2020Symmetry} are the basis for obtaining exact models of gravitational waves in this paper. These spaces allow the existence of "privileged" coordinate systems, where the Hamilton-Jacobi equation of test particles allows exact integration by the method of separation of variables, and one of the variables on which the space metric depends in privileged coordinate systems is the wave variable, along which the space-time interval goes to zero. At present, from the observational data of gravitational wave detection, it has been established with high accuracy that gravitational waves propagate at the speed of light \cite{Abbott2017PRL161101}.

Shapovalov spacetimes allow the construction of the complete integral of the Hamilton-Jacobi equation of test particles as a function of coordinates and a set of independent constant parameters determined by the initial or boundary conditions for the motion of test particles.
The possibility of constructing such a complete integral leads to many useful consequences: obtaining the exact form of test particle trajectories (i.e., the ability to find the explicit form of geodesics), the possibility of exact integrating the geodesic deviation equation and calculating tidal accelerations in a gravitational wave. All these possibilities make it possible to analytically calculate the secondary physical effects of a gravitational wave when interacting with other physical objects and fields~\cite{OsetrinDust2016,OsetrinRadiation2017,Obukhov202284,Obukhov2021695,%
Obukhov2022632,Obukhov2022142,Obukhov2021134,Obukhov2021183}, as well as in theories of gravity with quantum and other modifications~\cite{Odintsov2007,Odintsov2011,Capozziello2011,Odintsov2017}.

In addition, we recall that an important feature of the Shapovalov wave models under consideration is the possibility of making an explicit transition from privileged coordinate systems with wave variables to synchronous laboratory coordinate systems, where time and spatial coordinates are separated, and a freely falling observer is at rest, and time synchronization is also possible in different points of space \cite{LandauEng1}.

Useful properties of Shapovalov's wave spacetimes are based on their special symmetries, namely on the fact that they allow the existence of the so-called ''complete'' set of commuting vectors and Killing tensors of the second rank, which determine the presence of integrals of motion 
of test particles~\cite{Shapovalov1978I, Shapovalov1978II, Shapovalov1979} (independent constant parameters in the comlete integral of the Hamilton-Jacobi equation). The type of Shapovalov wave spaces is determined by the number of commuting Killing vectors included in the ''complete'' set (from 1 to 3 Killing vectors in the ''complete'' set), i.e., for example, the Shapovalov spaces of type III considered in the paper allow three Killing vectors in a ''complete'' set.


\section{
Shapovalov spacetimes and geodesic deviation
}

%
Let us recall the necessary information from the Hamilton-Jacobi formalism and from the theory of Shapovalov wave spacetimes for completeness.

Hamilton-Jacobi equation of a test particle in a gravitational field
has the form (see \cite{LandauEng1}):
\begin{equation}
g^{{\alpha}{\beta}}\frac{\partial S}{\partial x^{\alpha}}\frac{\partial S}{\partial x^{\beta}}=m^2c^2
,
\qquad
{\alpha},{\beta},{\gamma},{\delta}=0,...(n-1),
\label{HJE}
\end{equation}
where $m$ is the mass of the test particle, the capital letter $S$ denotes the action function of the test particle, $n$ is the dimension of space, $c$ is the speed of light, which we further set equal to unity.
\begin{definition}[Shapovalov spaces]
\label{Shapovalov_space}
If the space allows the existence of a privileged coordinate system $\{x^{\alpha}\}$, where the Hamilton-Jacobi equation (\ref{HJE}) can be integrated by the complete separation of variables method, when the complete integral $S$ for the test particle action function can be represented in a ''separated'' form:
\begin{equation} 
S=\phi_0(x^0,\lambda_0,...\lambda_{n-1})
+\phi_1(x^1,\lambda_0,...\lambda_{n-1})
+...+\phi_{n-1}(x^{n-1},\lambda_0,...\lambda_{n-1})
,
\end{equation} 
$$
\lambda_0, \lambda_1,...,\lambda_{n-1} - \mbox{const},
\qquad
\det \left\lvert\frac{\partial{}^2 S}{\partial x^{\alpha}\partial \lambda_{\beta}}\right\rvert \ne 
0
,
$$
moreover, one of the non-ignored variables (on which the metric depends) is a wave (null), i.e. the spacetime interval along this variable vanishes,
then such a space will be called the Shapovalov wave spacetime.
\end{definition}
Recall also that spacetimes that allow complete separation of variables in the Hamilton-Jacobi equation (\ref{HJE}) were first considered by Paul Stäckel \cite{Stackel1897145}, and
theory of these spaces
acquired a complete form in the works of Vladimir Shapovalov (see \cite{Shapovalov1978I,Shapovalov1978II, Shapovalov1979}), where a complete classification of such spaces was presented and an explicit form of the metrics of all given spaces in privileged coordinate systems was given. Shapovalov's classification also included wave spacetimes that allowed non-ignorable wave variables on which the metric depends in privileged coordinate systems.

The ability to construct the complete integral of the Hamilton-Jacobi equation in these spaces, in turn, allows us to find the trajectories of the test particles by determining the dependence of the coordinates $x^{\alpha}$ on the proper time $\tau$ of the test particle on the geodesic line of the particle by the following equations:
\begin{equation} 
\label{MovEqu}
\frac{\partial S (x^{\gamma},\lambda_{\alpha})}{\partial \lambda_{\beta}}=\sigma_{\beta},
\qquad
\tau=S (x^{\beta},\lambda_{\alpha})\Bigr\rvert_{m=1},
\end{equation} 
where $\lambda_\alpha$, $\sigma_\beta$ are additional independent constant parameters determined by the initial conditions of the test particle motion, $\tau$ is the proper time of the particle.


The resulting complete integral of the Hamilton-Jacobi equation of test particles also allows finding solutions for the geodesic deviation equations in the considered models of gravitational waves.
Recall that the deviation of geodesics is the basic manifestation of the gravitational field and the detection of gravitational waves. 
%

The geodesic deviation equation has the following form (see f.e. \cite{LandauEng1}):
\begin{equation}
\label{Deviation}
\frac{D^2 \eta^{\alpha}}{{d\tau}^2}=
 R^{\alpha}{}_{{\beta}{\gamma}{\delta}}u^{\beta} u^{\gamma}\eta^{\delta} ,
\end{equation}
where
$R^{\alpha}{}_{{\beta}{\gamma}{\delta}}$ is the Riemann curvature tensor of spacetime,
$u^{\alpha}$ is the 4-velocity of the test particle on the base geodesic line,
$\eta^{\alpha}$ is the geodesic deviation vector,
$D$ is the covariant derivative.
The coordinates $x^{\alpha}$ are parametrized by the proper time $\tau$ along the base geodesic line.

The 4-velocity of a particle is known to satisfy the condition \cite{LandauEng1}:
\begin{equation}
\label{Norm}
g^{{\alpha}{\beta}}u_{\alpha}u_{\beta}=1.
\end{equation}

The 4-velocity of a particle is related to its action function $S$ by the relation:
\begin{equation}
u_{\alpha}=\frac{\partial S}{\partial x^{\alpha}}\biggr\rvert_{m=1}.
\end{equation}

Thus, for Shapovalov wave spacetimes, the 4-velocity of a test particle $u^\alpha$ depends on $(n-1)$ parameters $\lambda_1,...,\lambda_{(n-1)}$:
\begin{equation} 
u_\alpha=u_\alpha(x^\beta,\lambda_1,...,\lambda_{(n-1)})
.
\end{equation} 
The proper time of a test particle can be represented in the following form:
\begin{equation} 
\tau =S(x^\alpha,\lambda_1,...,\lambda_{(n-1)})\Bigr\rvert_{m=1}
.\end{equation}


%
If the complete integral of the Hamilton-Jacobi equation of the test particle is found, then it is possible to obtain the exact solution for the geodesic deviation vector by the method found by variational methods in \cite{Bazanski19891018} as a solution to the system of equations of the following form:
\begin{equation} 
\label{Deviation1}
\eta^\alpha\,
\frac{\partial u_\alpha(x^\beta,\lambda_i)}{\partial\lambda_k} 
+\rho_i\,\frac{\partial^2 S(x^\alpha,\lambda_j)}{\partial\lambda_i\partial\lambda_k}=\vartheta_k
,
\end{equation}  
\begin{equation} 
\label{Deviation2}
u_\beta(x^\alpha,\lambda_k)\,\eta^\beta=0
,
\quad
\alpha,\beta,\gamma = 0\ldots 3;
\quad
i,j,k=1\ldots 3,
\end{equation} 
where $\lambda_k$, $\rho_k$, $\vartheta_k$ are independent constant parameters.


The constants $\lambda_k$ are given by the initial data for the test particle velocity on the base geodesic, and the constants $\rho_k$ and $\vartheta_k$ are given by the initial data on the adjacent geodesic. The equations of motion (\ref{MovEqu}) define the dependence of coordinates on proper time $\tau$.

\section{Type III Shapovalov wave spacetimes}

Type III Shapovalov wave spaces allow the existence of three commuting Killing vectors, so the metric of these spaces in the privileged coordinate system can generally be written in the following form \cite{OsetrinHomog312002,OsetrinHomog2006,OsetrinHomog312020}:
\begin{equation} 
{\rm g}^{\alpha\beta}(x^0) = \left(                                                      
              \begin{array}{cccc}
               0 & 1 & {g^{02}}(x^0)  & {g^{03}}(x^0)   \\
               1 & 0 & 0 & 0 \\
               {g^{02}}(x^0)   & 0 & {g^{22}}(x^0)  & {g^{23}}(x^0)  \\
               {g^{03}}(x^0)  & 0 & {g^{23}}(x^0)  & {g^{33}}(x^0)  \\
              \end{array}
              \right)
\label{MetricShapovalovIII}
,\end{equation} 
where indices $\alpha$, $\beta$ run through the values 0,1,2,3.
The variable $x^0$ is a wave (null) variable along which the spacetime interval vanishes.
Thus, the metric is defined in the general case by five functions of the wave variable $x^0$.


The spacetime for a gravitational wave (\ref{MetricShapovalovIII}) has type N according to Petrov's classification.


Einstein's equations with cosmological constant $\Lambda$ in vacuum
\begin{equation}
R_{\alpha\beta}=\Lambda g_{\alpha\beta}
\label{EinsteinEqs}
\end{equation}
for the metric (\ref{MetricShapovalovIII}) lead to the following necessary conditions:
\begin{equation}
\Lambda=g^{02}=g^{03}=0
\label{LambdaFromEinsteinEqs}
.\end{equation}

Thus, the metric of Shapovalov spacetimes of type III, taking into account the restrictions of Einstein's vacuum equations, will take the following form in the privileged coordinate system:
\begin{equation} 
{\rm g}^{\alpha\beta}(x^0) = \left(                                                      
              \begin{array}{cccc}
               0 & 1 & 0 & 0 \\
               1 & 0 & 0 & 0 \\
               0 & 0 & {g^{22}}(x^0)  & {g^{23}}(x^0)  \\
               0 & 0 & {g^{23}}(x^0)  & {g^{33}} (x^0) \\
              \end{array}
              \right)
,\quad
{g^{22}} {g^{33}}-({g^{23}})^2>0
\label{PrivilMetricUp}
.\end{equation} 
The space under consideration is a plane-wave spacetime and admits a covariantly constant vector that specifies the direction of propagation of a gravitational wave:
\begin{equation} 
\nabla_\alpha K_\beta=0
\quad
\to
\quad
K^\alpha=\bigl( 0,1,0,0 \bigr)
.
\end{equation} 

In addition to the relations (\ref{LambdaFromEinsteinEqs}), vacuum Einstein's equations contain equation $R_{00}=0$, which allows one to express the second derivative of one of the functions in the metric. Expressing ${g^{33}}''(x^0)$ from the remaining equation, we obtain the relation of the following form:
$$
{g^{33}}''=
\frac{
6 ({g^{23}})^2 \left({g^{22}}' {g^{33}}' + ({g^{23}}'{})^2\right)
-12 {g^{22}} {g^{23}} {g^{23}}' {g^{33}}' + 3 ({g^{22}})^2 ({g^{33}}'{})^2
}{2 {g^{22}} \left({g^{22}} {g^{33}} - ({g^{23}})^2\right)}
$$
$$
\mbox{}
+
 {g^{33}} 
 \,
\frac{
 {g^{22}}'' ({g^{23}})^2+3 {g^{22}} ({g^{23}}'{})^2
+{g^{23}} \left(2 {g^{23}}'' {g^{22}} - 6 {g^{22}}' {g^{23}}' \right)
}{ {g^{22}} \left({g^{22}} {g^{33}} - ({g^{23}})^2\right)}
$$
\begin{equation} 
\mbox{}
+
\frac{
({g^{33}})^2 \left(3   ({g^{22}}'{})^2-2 {g^{22}}'' {g^{22}} \right)
-4 {g^{23}}'' ({g^{23}})^3
}{2 {g^{22}} \left({g^{22}} {g^{33}} - ({g^{23}})^2\right)}
\label{EqR00}
,\end{equation} 
%
%
where the top prime denotes the derivative of the function with respect to the variable $x^0$.

Thus, the metric has three functions ${g^{22}}(x^0)$, ${g^{23}}(x^0)$ and ${g^{33}}(x^0) $, connected by one ordinary differential equation (\ref{EqR00}), which arises from the system of Einstein's field equations in vacuum (the cosmological constant $\Lambda$ vanishes).

The test particle action function in the privileged coordinate system takes the ''separated'' form:
\begin{equation} 
S={\phi_0}({x^0}) 
+
\sum_{k=1}^3 {\lambda_k} x^k
,\end{equation} 
%
where $\lambda_k$ are constant patameters determined by the initial or boundary values of the velocities (momentums) of the test particle.

The equations of motion of test particles in the Hamilton-Jacobi formalism can be integrated in the privileged coordinate system using the following notation for integrals of metric functions (\ref{PrivilMetricUp}):
\begin{equation} 
{\Theta_{ab}}(x^0)=\int{g^{ab}(x^0)}\,dx^0
,\qquad
a,b,c=2,3
\label{IntegralsOfMetric}
.\end{equation} 

The Hamilton-Jacobi equation (\ref{HJE}) can be integrated and we get the following form of the function $\phi_0 (x^0)$ (the test particle mass $m$ is set equal to unity):
\begin{equation} 
 \phi_0 (x^0) = -\frac{{\lambda_2}^2 {\Theta_{22}} +2 {\lambda_2} {\lambda_3} {\Theta_{23}} +{\lambda_3}^2 {\Theta_{33}} -{x^0}}{2 {\lambda_1}} 
.\end{equation} 
%
Thus, we have obtained the explicit form of the complete integral $S(x^\alpha, \lambda_k)$ for the Hamilton-Jacobi equation for a test particle.

Then, in the Hamilton-Jacobt formalism, the trajectories of test particles
(\ref{MovEqu}) in the privileged coordinate system is found in the following form:
\begin{equation} 
\label{PrivX0}
x^0 (\tau)  = {\lambda_1} {\tau} 
,\end{equation} 
\begin{equation} 
x^1 (\tau)   = 
\frac{{\tau}}{2{\lambda_1}}
-\frac{{\lambda_2}^2 {\Theta_{22}} + 2 {\lambda_2} {\lambda_3} {\Theta_{23}} + {\lambda_3}^2 {\Theta_{33}} 
}{2 {\lambda_1}^2} 
\biggr\rvert_{x^0   = {\lambda_1} {\tau} }
\label{PrivX1}
,\end{equation} 
\begin{equation} 
x^2  (\tau)  = \frac{{\lambda_2} {\Theta_{22}} + {\lambda_3} {\Theta_{23}}}{{\lambda_1}} 
\biggr\rvert_{x^0   = {\lambda_1} {\tau} }
\label{PrivX2}
,\end{equation} 
\begin{equation} 
x^3 (\tau)   = \frac{{\lambda_2} {\Theta_{23}} + {\lambda_3} {\Theta_{33}}}{{\lambda_1}}
\biggr\rvert_{x^0   = {\lambda_1} {\tau} } 
\label{PrivX3}
,\end{equation} 
%
where $\tau$ is the proper time of the test particle, the parameters ${\lambda_k}$ are determined by the initial (boundary) values of the velocities (momentums) of the test particle. The constants $\sigma_k$ are set equal to zero by choosing the origin of the coordinates.

For the components of the 4-velocity of the test particle $u^\alpha$ in the privileged coordinate system, from the equations of the trajectory we obtain the following expressions:
\begin{equation} 
u^{0} = {\lambda_1} 
,\end{equation} 
\begin{equation} 
u^{1} (\tau)  = -\frac{{\lambda_2}^2 {g^{22}} +2 {\lambda_2} {\lambda_3} {g^{23}} +{\lambda_3}^2 {g^{33}} -1}{2 {\lambda_1}} 
\biggr\rvert_{x^0   = {\lambda_1} {\tau} } 
,\end{equation} 
\begin{equation} 
u^{2} (\tau)  =\left( {\lambda_2} {g^{22}} +{\lambda_3} {g^{23}} \right) 
\Bigr\rvert_{x^0   = {\lambda_1} {\tau} } 
,\end{equation} 
\begin{equation} 
u^{3} (\tau)  =\left(  {\lambda_2} {g^{23}} +{\lambda_3} {g^{33}} \right) 
\Bigr\rvert_{x^0   = {\lambda_1} {\tau} } 
. \end{equation} 

Now we have all the necessary relations in order to write down and solve equations for the geodesic deviation vector (\ref{Deviation1})-(\ref{Deviation2}) for the metric (\ref{PrivilMetricUp}).
Omitting obvious calculations, we present the solution of the system of equations
for the geodesic deviation vector $ \eta^\alpha (\tau)$ in the privileged coordinate system:
\begin{equation}
\label{PrivilCommonEta0}
 \eta^0 (\tau) = 
{\rho_1}{\tau}
 -{\lambda_1} {\Omega} 
 ,\qquad
 x^0 = {\lambda_1} {\tau} 
, \end{equation}
$$
 \eta^1 (\tau)  = 
 -\frac{
 \left(
 {\rho_1} {\tau}
 -{\lambda_1} {\Omega} 
 \right)
 \left(
 {\lambda_2}^2 {g^{22}} +2 {\lambda_2} {\lambda_3} {g^{23}} +{\lambda_3}^2 {g^{33}} 
 +1
 \right)
}{2 {\lambda_1}^2}
$$
\begin{equation}
\mbox{}
 -\frac{
{\Theta_{23}}{} ({\lambda_2} {R_3}+{\lambda_3} {R_2})
+{\lambda_2} {R_2} {\Theta_{22}} 
+ {\lambda_3} {R_3} {\Theta_{33}}}{ {\lambda_1}^3}
-\frac{\Omega}{{\lambda_1}}+{\vartheta_1}
\label{PrivilCommonEta1}
, \end{equation}
\begin{equation}
\label{PrivilCommonEta2}
 \eta^2 (\tau)  = 
 \frac{
 {\lambda_1}
 \left(
 {\rho_1} {\tau}
 -{\lambda_1} {\Omega} 
 \right) 
 \left({\lambda_2} {g^{22}} +{\lambda_3} {g^{23}} \right)
 +{R_2} {\Theta_{22}} +{R_3} {\Theta_{23}}
 }{{\lambda_1}^2}
 +{\vartheta_2} 
, \end{equation}
\begin{equation}
\label{PrivilCommonEta3}
 \eta^3 (\tau)  = \frac{
  {\lambda_1}
 \left(
 {\rho_1} {\tau}
 -{\lambda_1} {\Omega} 
 \right)
 \left({\lambda_2} {g^{23}} +{\lambda_3} {g^{33}} \right)+{R_2} {\Theta_{23}} +{R_3} {\Theta_{33}}
 }{{\lambda_1}^2}
 +{\vartheta_3} 
, \end{equation}
%
where $\tau$ is the proper time of the test particle on the base geodesic line. The functions $g^{ab}$ and ${\Theta_{ab}}$ (\ref{IntegralsOfMetric}) here are functions of one variable $x^0$, which on a geodesic with proper time $\tau$ has the form $x ^0 = {\lambda_1}{\tau}$.

The formulas (\ref{PrivilCommonEta0})-(\ref{PrivilCommonEta3}) include both independent parameters $\lambda_k$, ${\rho_k}$, and ${\vartheta_k}$, as well as dependent auxiliary constants introduced to shorten the notation:
\begin{equation} 
R_2 = {\lambda_1} {\rho_2}-{\lambda_2} {\rho_1}
,\qquad
R_3 = {\lambda_1} {\rho_3}-{\lambda_3} {\rho_1}
\label{R2R3}
,\end{equation} 
\begin{equation} 
\Omega = {\lambda_1} {\vartheta_1}+{\lambda_2} {\vartheta_2}+{\lambda_3} {\vartheta_3}
\label{Omega}
,\end{equation} 
%
%

The the constant parameters $\lambda_k$ are determined by the initial or boundary values for the momenta of a test particle on the base geodesic, and the parameters $\rho_k$ and ${\vartheta_k}$ are determined by the initial or boundary values of the momenta and the relative position of particles on neighboring geodesics.

The resulting deviation vector now allows us to calculate the deviation velocity $D\eta^\alpha/d\tau$ and the tidal acceleration $D^2\eta^\alpha/{d\tau}^2$.

Let us find the deviation rate $V^\alpha (\tau)=D\eta^\alpha/d\tau$ for the resulting general solution of the deviation equation (\ref{PrivilCommonEta0})-(\ref{PrivilCommonEta3}) in the metric (\ref {PrivilMetricUp}) in the privileged coordinate system:
\begin{equation}
\label{PrivCommonV0}
V^0 =   {\rho_1} 
,\end{equation} 
$$
V^1 (\tau) = \frac{1}{ 2 {\lambda_1}^2 \left(({g^{23}})^2-{g^{22}} {g^{33}} \right) }
\biggl[ 
-{g^{33}}{} ({R_2} {\Theta_{22}} +{R_3} {\Theta_{23}}{}) \left({\lambda_2} {g^{22}}' +{\lambda_3} {g^{23}}' \right)
$$ $$
{\lambda_1}^2 {\vartheta_2} \left({g^{23}} \left({\lambda_2} {g^{23}}' +{\lambda_3} {g^{33}}' \right)-{g^{33}} \left({\lambda_2} {g^{22}}' +{\lambda_3} {g^{23}}' \right)\right)
$$ $$
+{\lambda_1}^2 {\vartheta_3} \left({g^{23}} \left({\lambda_2} {g^{22}}' +{\lambda_3} {g^{23}}' \right)-{g^{22}} \left({\lambda_2} {g^{23}}' +{\lambda_3} {g^{33}}' \right)\right)
$$ $$
+{\rho_1} \left({g^{22}} {g^{33}} -({g^{23}})^2\right) \left({\lambda_2}^2 {g^{22}} +2 {\lambda_2} {\lambda_3} {g^{23}} +{\lambda_3}^2 {g^{33}} +1\right)
$$
$$
+{g^{23}} 
\biggl(
{\Theta_{23}} \left({g^{23}}'{} ({\lambda_2} {R_3}+{\lambda_3} {R_2})+{\lambda_2} {R_2} {g^{22}}' +{\lambda_3} {R_3} {g^{33}}' \right)
$$ $$
+{R_3} {\Theta_{33}} \left({\lambda_2} {g^{22}}' +{\lambda_3} {g^{23}}' \right)+{R_2} {\Theta_{22}} \left({\lambda_2} {g^{23}}' +{\lambda_3} {g^{33}}' \right)
\biggr)
$$ $$
-{g^{22}} 
\biggl(
({R_2} {\Theta_{23}} +{R_3} {\Theta_{33}}{}) \left({\lambda_2} {g^{23}}' +{\lambda_3} {g^{33}}' \right)
$$ $$
-2 {g^{23}} {g^{33}}{} ({\lambda_2} {R_3}+{\lambda_3} {R_2})+2 {\lambda_2} {R_2} ({g^{23}})^2-2 {\lambda_3} {R_3} {g^{33}}^2
\biggr)
$$ 
\begin{equation}
-2 ({g^{23}})^3 ({\lambda_2} {R_3}+{\lambda_3} {R_2})+2 {\lambda_2} {R_2} {g^{22}}^2 {g^{33}} -2 {\lambda_3} {R_3} ({g^{23}})^2 {g^{33}} 
\biggr]
\label{PrivCommonV1}
,\end{equation}
$$
V^2 (\tau)  =\frac{1}{2 {\lambda_1} \left(({g^{23}})^2-{g^{22}} {g^{33}} \right) }
\biggl[
 {\lambda_1}^2 {\vartheta_2} \left({g^{22}}' {g^{33}} -{g^{23}} {g^{23}}' \right)
$$ $$
 +{\lambda_1}^2 {\vartheta_3} \left({g^{22}} {g^{23}}' -{g^{22}}' {g^{23}} \right)-2 {\rho_1} \left({g^{22}} {g^{33}} 
 -({g^{23}})^2\right) \left({\lambda_2} {g^{22}} +{\lambda_3} {g^{23}} \right)
 $$ $$
 -{g^{23}} \left({g^{22}}'({R_2} {\Theta_{23}} +{R_3} {\Theta_{33}} )
 +{g^{23}}'({R_2} {\Theta_{22}}  +{R_3} {\Theta_{23}} ) \right)
 $$ $$
  +{g^{22}} \left({g^{23}}'{} ({R_2} {\Theta_{23}} +{R_3} {\Theta_{33}}{})-2 {R_3} {g^{23}} {g^{33}} +2 {R_2} ({g^{23}})^2\right)
  $$ 
\begin{equation}
  +{g^{22}}' {g^{33}}{} ({R_2} {\Theta_{22}} +{R_3} {\Theta_{23}}{})
 -2 {R_2} {g^{22}}^2 {g^{33}} +2 {R_3} {g^{23}}^3 
\label{PrivCommonV2}
\biggr]
,\end{equation}
$$
V^3  (\tau)  = \frac{1}{ 2 {\lambda_1} \left(({g^{23}})^2-{g^{22}} {g^{33}} \right)}
\biggl[
 {\lambda_1}^2 {\vartheta_3} \left({g^{22}} {g^{33}}' -{g^{23}} {g^{23}}' \right)
 $$ $$
 +{\lambda_1}^2 {\vartheta_2} \left({g^{23}}' {g^{33}} -{g^{23}} {g^{33}}' \right)
 +2 {\rho_1} \left(({g^{23}})^2-{g^{22}} {g^{33}} \right) \left({\lambda_2} {g^{23}} +{\lambda_3} {g^{33}} \right)
 $$ $$
 -{g^{23}} \left(2 {R_2} {g^{22}} {g^{33}} +{R_2} {\Theta_{22}} {g^{33}}' +{R_3} {\Theta_{33}} {g^{23}}' +{R_2} {\Theta_{23}} {g^{23}}' +{R_3} {\Theta_{23}} {g^{33}}' \right)
 $$ $$
 +{g^{22}} \left({g^{33}}'{} ({R_2} {\Theta_{23}} +{R_3} {\Theta_{33}}{})-2 {R_3} {g^{33}}^2\right)
 $$ 
 \begin{equation}
 +{g^{23}}' {g^{33}}{} ({R_2} {\Theta_{22}} +{R_3} {\Theta_{23}}{})+2 {R_3} ({g^{23}})^2 {g^{33}} +2 {R_2} {g^{23}}^3 
 \biggr]
 \label{PrivCommonV3}
.\end{equation}

Recall that the functions $g^{ab}$ and $\Theta_{ab}$ (\ref{IntegralsOfMetric}) depend on one variable $x^0$, which is equal to $x^0=\lambda_1\tau$ on the base geodesic, where $\tau$ is proper time on the base geodesic. The constants $\lambda_k$ are determined by the initial or boundary conditions for the velocity on the base geodesic, the constants $\rho_k$ and $\vartheta_k$ are determined by the initial or boundary conditions for the velocity and relative position on the adjacent geodesic.
The constants $R_a$ and $\Omega$ are defined by the relations (\ref{R2R3})-(\ref{Omega}).

The expressions for the tidal acceleration $A^\alpha=D^2\eta^\alpha/{d\tau}^2$ have a cumbersome form and we have moved them to the appendix (\ref{AppendixA}).

Trajectories of test particles $x^\alpha(\tau)$ (\ref{PrivX0})-(\ref{PrivX3}),
geodesic deviation vector $\eta^\alpha(\tau)$ (\ref{PrivilCommonEta0})-(\ref{PrivilCommonEta3}) and geodesic deviation rate $V^\alpha(\tau)$ (\ref{PrivCommonV0})-(\ref{PrivCommonV3}), found above for Shapovalov type III gravitational waves in general form, i.e. obtained without taking into account the Einstein equation (\ref{EqR00}). Therefore, they can be applied to the metric (\ref{PrivilMetricUp}) in any field equations based on it, including application in modified gravity theories (see \cite{Odintsov2007,Odintsov2011,Capozziello2011,Odintsov2017}).

The obtained characteristics of a gravitational wave, in turn, make it possible to evaluate the influence of a gravitational wave on physical objects, fields, and on the physical media through which it passes. Including calculate the radiation of charges moving with tidal acceleration in a gravitational wave.

\section{Synchronous reference system
}

%
The use of a privileged coordinate system allowed us to integrate the equations of motion of test particles and obtain an exact solution for the geodesic deviation vector, obtain an exact form of the deviation velocity and tidal accelerations in a gravitational wave.

The form of trajectories of test particles in the field of a gravitational wave (\ref{PrivX0})-(\ref{PrivX3}) obtained using the complete integral of the Hamilton-Jacobi equation makes it possible to additionally implement
transition from the privileged coordinate system used to the laboratory synchronous reference frame ${\tilde x}{}^k$ associated with an observer freely falling along the base geodesic.
The advantages of a synchronous reference system are related to the fact that time and spatial variables in this reference system are separated, and time at different points in space can be synchronized.

The transition to the synchronous coordinate system ${\tilde x}{}^k$ can be explicitly implemented using the relations (\ref{PrivX0})-(\ref{PrivX3}) according to the rules (see \cite{LandauEng1}):
\begin{equation} 
x^\alpha \to {\tilde x}{}^\alpha=\left(\tau,\lambda_1,\lambda_2,\lambda_3 \right)
.\end{equation} 

Thus, transformations from the privileged coordinate system $\{x^\alpha\}$ to the synchronous reference system $\{{\tilde x{}^\alpha}\}$ take the following form:
\begin{equation} 
\label{TransitionX0}
x^0 = {\tilde x{}^1} {\tau} 
, \end{equation} 
\begin{equation} 
\label{TransitionX1}
x^1 = -\frac{({\tilde x{}^2})^2 {\Theta_{22}} + 2 {\tilde x{}^2} {\tilde x{}^3} {\Theta_{23}} + ({\tilde x{}^3})^2 {\Theta_{33}} - {\tilde x{}^1} {\tau}}{2 ({\tilde x{}^1})^2} 
\biggr\rvert_{x^0   = {\tilde x{}^1} {\tau} } 
, \end{equation} 
\begin{equation} 
\label{TransitionX2}
x^2 = \frac{{\tilde x{}^2} {\Theta_{22}} + {\tilde x{}^3} {\Theta_{23}}}{{\tilde x{}^1}} 
\biggr\rvert_{x^0   = {\tilde x{}^1} {\tau} } 
, \end{equation} 
\begin{equation} 
x^3 = \frac{{\tilde x{}^2} {\Theta_{23}} + {\tilde x{}^3} {\Theta_{33}}}{{\tilde x{}^1}} 
\biggr\rvert_{x^0   = {\tilde x{}^1} {\tau} } 
\label{TransitionX3}
. \end{equation} 

In the resulting synchronous frame of reference, the 4-velocity of the test particle will take the form
$
\tilde u{}^k=\left\{ 1,\, 0,\, 0,\, 0\right\}
$.
Thus, the observer rests on the base geodesic in the obtained synchronous frame of reference, and the observer's proper time $\tau$ becomes the time of the frame of reference. The metric (\ref{PrivilMetricUp}) in the synchronous reference system will take the following form:
\begin{equation}
{ds}^2={d\tau}^2-{dl}^2={d\tau}^2+\tilde g{}_{ij}\bigl(\tau,{\tilde x{}^ k}\bigr){d\tilde x{}^i}{d\tilde x{}^j}
, \end{equation}
where $dl$ is the spatial distance element, and $\tau$ is the time in the synchronous frame of reference (the speed of light $c$ is set equal to unity).

When switching to a synchronous frame of reference, it is required in the functions $g^{ab}(x^0)$ and $\Theta_{ab}(x^0)$ to change
 $
 x^0 \, \to \, \tau {\tilde x{}^1}
 ,$
in this case, the components of the gravitational wave metric (\ref{PrivilMetricUp}) will take the following form in the synchronous reference frame $\{{\tilde x{}^\alpha}\}$:
\begin{equation}
\tilde g{}_{00} = 1 
,\qquad
\tilde g{}_{0k} = 0
\label{metricSynchr0k}
, \end{equation}
$$
\tilde g{}_{11}\bigl(\tau,{\tilde x{}^k}\bigr) = 
-\frac{
{\tau}^2 
}{({\tilde x{}^1})^2 
} 
-\frac{
{\Theta_{23}}^2 
\left(({\tilde x{}^2})^2 {g^{22}} + {\tilde x{}^3} \left({\tilde x{}^3} {g^{33}} - 2 {\tilde x{}^2} {g^{23}} \right)\right)
}{({\tilde x{}^1})^4 \left( ({g^{23}})^2-{g^{22}} {g^{33}} \right)} 
$$
$$
+\frac{
2 {\Theta_{23}} \left({\tilde x{}^3} {\Theta_{33}} \left({\tilde x{}^3} {g^{23}} - {\tilde x{}^2} {g^{22}} \right)
+{\tilde x{}^2} {\Theta_{22}} \left({\tilde x{}^2} {g^{23}} - {\tilde x{}^3} {g^{33}} \right)\right)
}{({\tilde x{}^1})^4 \left( ({g^{23}})^2-{g^{22}} {g^{33}} \right)} 
$$
\begin{equation}
-\frac{
({\tilde x{}^2})^2 {\Theta_{22}}^2 {g^{33}} 
-2 {\tilde x{}^2} {\tilde x{}^3} {\Theta_{22}} {\Theta_{33}} {g^{23}} + ({\tilde x{}^3})^2 {\Theta_{33}}^2 {g^{22}}
}{({\tilde x{}^1})^4 \left( ({g^{23}})^2-{g^{22}} {g^{33}} \right)} 
, \end{equation}
$$
\tilde g{}_{12} \bigl(\tau,{\tilde x{}^k}\bigr)= \frac{{\tilde x{}^2} \left({\Theta_{23}}^2 {g^{22}} - 2 {\Theta_{22}} {\Theta_{23}} {g^{23}} + {\Theta_{22}}^2 {g^{33}} \right)}{({\tilde x{}^1})^3 \left( ({g^{23}})^2-{g^{22}} {g^{33}} \right)}
$$ 
\begin{equation}
+\frac{{\tilde x{}^3} \left({\Theta_{23}} {\Theta_{33}} {g^{22}} - {\Theta_{22}} {\Theta_{33}} {g^{23}} + {\Theta_{22}} {\Theta_{23}} {g^{33}} - {\Theta_{23}}^2 {g^{23}} \right)
}{({\tilde x{}^1})^3 \left(({g^{23}})^2-{g^{22}} {g^{33}} \right)} 
, \end{equation}
$$
\tilde g{}_{13}\bigl(\tau,{\tilde x{}^k}\bigr) = \frac{{\tilde x{}^2} \left({\Theta_{23}} {\Theta_{33}} {g^{22}} - {\Theta_{22}} {\Theta_{33}} {g^{23}} + {\Theta_{22}} {\Theta_{23}} {g^{33}}{}-{\Theta_{23}}^2 {g^{23}} \right)}{({\tilde x{}^1})^3 \left(({g^{23}})^2-{g^{22}} {g^{33}} \right)}
$$
\begin{equation}
+\frac{{\tilde x{}^3} \left({\Theta_{33}}^2 {g^{22}} - 2 {\Theta_{23}} {\Theta_{33}} {g^{23}} + {\Theta_{23}}^2 {g^{33}} \right)}{({\tilde x{}^1})^3 \left(({g^{23}})^2-{g^{22}} {g^{33}} \right)} 
, \end{equation}
\begin{equation}
\tilde g{}_{22}\bigl(\tau,{\tilde x{}^k}\bigr) = -\frac{{\Theta_{23}}^2 {g^{22}} - 2 {\Theta_{22}} {\Theta_{23}} {g^{23}}{}+{\Theta_{22}}^2 {g^{33}}}{({\tilde x{}^1})^2 \left(({g^{23}})^2-{g^{22}} {g^{33}} \right)} 
, \end{equation}
\begin{equation}
\tilde g{}_{23}\bigl(\tau,{\tilde x{}^k}\bigr) = 
\frac{
{\Theta_{22}} {\Theta_{33}} {g^{23}} 
+ {\Theta_{23}}^2 {g^{23}}
-{\Theta_{23}} \left({\Theta_{33}} {g^{22}} + {\Theta_{22}} {g^{33}}{}\right)
}{({\tilde x{}^1})^2 \left(({g^{23}})^2-{g^{22}} {g^{33}} \right)} 
, \end{equation}
\begin{equation}
\tilde g{}_{33}\bigl(\tau,{\tilde x{}^k}\bigr) = -\frac{{\Theta_{33}}^2 {g^{22}} - 2 {\Theta_{23}} {\Theta_{33}} {g^{23}}{}+{\Theta_{23}}^2 {g^{33}}}{({\tilde x{}^1})^2 \left(({g^{23}})^2-{g^{22}} {g^{33}} \right)} 
\label{metricSynchr33}
, \end{equation}
%
where the one-variable functions $g^{ab}(x^0)$ and $\Theta_{ab}(x^0)$ (\ref{IntegralsOfMetric}) now depend on the product $\tau {\tilde x{}^ 1}$, and the observer's proper time on the base geodesic $\tau$ is the universal time in the synchronous frame of reference.

A metric with superscripts $\tilde g{}^{\alpha\beta}\bigl(\tau,{\tilde x{}^k}\bigr)$ has a more compact and clearer form in a synchronous frame of reference:
\begin{equation}
\tilde g{}^{00} = 1 
,\qquad
\tilde g{}^{0k} = 0 
\label{SynchrMetricUp0k}
, 
\qquad
\tilde g{}^{1k} \bigl(\tau,{\tilde x{}^i}\bigr)  = -\frac{{\tilde x{}^1} {\tilde x{}^k}}{{\tau}^2} 
, \end{equation}
\begin{equation} 
\tilde g{}^{ab} \bigl(\tau,{\tilde x{}^i}\bigr)  
= 
-\frac{{\tilde x{}^a} {\tilde x{}^b}}{{\tau}^2} 
+
({\tilde x{}^1})^2 F^{ab}(x)
\biggr\rvert_{x  = {\tilde x{}^1} {\tau} } 
\label{SynchrMetricUpab}
,\end{equation} 
where three independent functions of one variable $F^{ab}(x)$ are related to the metric functions in the privileged coordinate system by the following relations:
\begin{equation} 
F^{22}({\tilde x{}^1} {\tau})=
\frac{ 
{\Theta_{33}}^2 {g^{22}} - 2 {\Theta_{23}} {\Theta_{33}} {g^{23}} + {\Theta_{23}}^2 {g^{33}} 
}{\left({\Theta_{23}}^2-{\Theta_{22}} {\Theta_{33}} \right)^2}
\Biggr\rvert_{x^0  \to {\tilde x{}^1} {\tau} } 
\label{F22}
,\end{equation} 
\begin{equation} 
F^{23}=F^{32}=
\frac{ 
{\Theta_{22}} {\Theta_{33}} {g^{23}} 
-{\Theta_{23}} \left({\Theta_{33}} {g^{22}} + {\Theta_{22}} {g^{33}} 
-{\Theta_{23}} {g^{23}} 
\right)
}{\left({\Theta_{23}}^2-{\Theta_{22}} {\Theta_{33}} \right)^2}
\Biggr\rvert_{x^0  \to {\tilde x{}^1} {\tau} } 
\label{F23}
,\end{equation} 
\begin{equation} 
F^{33}({\tilde x{}^1} {\tau})=
\frac{
{\Theta_{23}}^2 {g^{22}} - 2 {\Theta_{22}} {\Theta_{23}} {g^{23}} + {\Theta_{22}}^2 {g^{33}} 
}{\left({\Theta_{23}}^2-{\Theta_{22}} {\Theta_{33}} \right)^2}
\Biggr\rvert_{x^0  \to {\tilde x{}^1} {\tau} } 
\label{F33}
.\end{equation} 

%
Thus, for calculations with a gravitational wave (\ref{PrivilMetricUp}) in a synchronous (laboratory) frame of reference associated with a freely falling observer, we can initially proceed from the general form of the metric
(\ref{SynchrMetricUp0k})-(\ref{SynchrMetricUpab})
 without using the relations (\ref{F22})-(\ref{F33}), defining three independent functions of one variable $F^{ab}({\tilde x{}^1} {\tau})$ in the synchronous system reference directly from the field equations of the corresponding theory of gravity.
 
Geodesic deviation vector $\tilde\eta{}^\alpha(\tau)$ in a gravitational wave with metric
 (\ref{SynchrMetricUp0k})-(\ref{SynchrMetricUpab})
after transformation of coordinates (\ref{TransitionX0})-(\ref{TransitionX1}) in the synchronous reference system will take the form:
\begin{equation}
\label{SynchrEta0}
\tilde\eta{}^0= 0 
, \end{equation}
\begin{equation}
\label{SynchrEta1}
\tilde\eta{}^1 (\tau) = {\rho_1}-\frac{{\lambda_1} {\Omega}}{{\tau}} 
, \end{equation}
\begin{equation}
\label{SynchrEta2}
\tilde\eta{}^2 (\tau) = 
{\rho_2}
-\frac{{\lambda_2} {\Omega}}{{\tau}} 
+\frac{{\lambda_1} {\vartheta_2} {\Theta_{33}}}{{\Theta_{22}} {\Theta_{33}} - {\Theta_{23}}^2}
+\frac{{\lambda_1} {\vartheta_3} {\Theta_{23}}}{{\Theta_{23}}^2-{\Theta_{22}} {\Theta_{33}}}
, \end{equation}
\begin{equation}
\label{SynchrEta3}
\tilde\eta{}^3 (\tau) = 
{\rho_3}
-\frac{{\lambda_3} {\Omega}}{{\tau}} 
+\frac{{\lambda_1} {\vartheta_2} {\Theta_{23}}}{{\Theta_{23}}^2-{\Theta_{22}} {\Theta_{33}}}
+\frac{{\lambda_1} {\vartheta_3} {\Theta_{22}}}{{\Theta_{22}} {\Theta_{33}} - {\Theta_{23}}^2}
, \end{equation}
%
where the constant parameters $\lambda_k$ and $\vartheta_k$ are determined by 
the initial (boundary) values of  the velocities on neighboring geodesics, the constants $\rho_k$ are determined by the initial (boundary) values of  the relative deviation of the geodesics. The auxiliary constant $\Omega$ is defined by the relation (\ref{Omega}). The functions $\Theta_{ab}(x^0)$ (\ref{IntegralsOfMetric}) depend on one variable equal to the product of $\lambda_1\tau$, where the variable $\tau$ is the time in the synchronous frame of reference.

Let us find the deviation rate $\tilde V{}^\alpha(\tau)=D\tilde\eta{}^\alpha/d\tau$ in the synchronous frame of reference using coordinate transformations (\ref{TransitionX0})-(\ref{TransitionX1}):
\begin{equation} 
\tilde V{}^0 = 0
,\end{equation} 
\begin{equation} 
\tilde V{}^1 
={\rho_1}/{\tau} 
,\end{equation} 
$$
\tilde V{}^2 
= 
\frac{1}{2 {\lambda_1} {\tau} \left({\Theta_{23}}^2-{\Theta_{22}} {\Theta_{33}}\right) \left(({g^{23}})^2-{g^{22}} {g^{33}} \right)}
\Biggl[
 {\lambda_1}^3 {\tau} {\vartheta_2} 
\biggl(
 {\Theta_{33}} \left({g^{23}} {g^{23}}' - {g^{22}}'  {g^{33}} \right)
$$ $$
+{\Theta_{23}} \left({g^{23}}' {g^{33}} -{g^{23}} {g^{33}}' \right)
\biggr)
$$ $$ 
 +{\lambda_1}^3 {\tau} {\vartheta_3} \left({\Theta_{23}} \left({g^{22}} {g^{33}}' - {g^{23}} {g^{23}}' \right)+{\Theta_{33}} \left({g^{22}}' {g^{23}} - {g^{22}} {g^{23}}'
  \right)\right)
 $$ $$
 +{\Theta_{23}}^2 
\biggl(
 {g^{22}} \left({\lambda_1} {\tau} {R_2} {g^{33}}' - 2 {\lambda_2} {\rho_1} {g^{33}} \right)
 - {\lambda_1} {\tau} {g^{23}} \left({R_2} {g^{23}}' + {R_3} {g^{33}}' \right)
$$ $$
  +2 {\lambda_2} {\rho_1} ({g^{23}})^2+{\lambda_1} {\tau} {R_3} {g^{23}}' {g^{33}} 
\biggr)
 $$ $$
 +{\Theta_{33}} 
\biggl(
 {\Theta_{22}} \left({g^{33}} \left(2 {\lambda_2} {\rho_1} {g^{22}} - {\lambda_1} {\tau} {R_2} {g^{22}}' \right)
 -2 {\lambda_2} {\rho_1} ({g^{23}})^2
 +{\lambda_1} {\tau} {R_2} {g^{23}} {g^{23}}' 
\right)
$$ $$
 +{\lambda_1} {\tau} 
\Bigl(
 {R_3} {\Theta_{33}}  {g^{22}}' {g^{23}} 
 - {g^{22}} 
\left(
{R_3} {\Theta_{33}} {g^{23}}' - 2 {R_3} {g^{23}} {g^{33}} + 2 {R_2} ({g^{23}})^2
\right)
$$ $$
+2 {R_2} {g^{22}}^2 {g^{33}} - 2 {R_3} {g^{23}}^3
\Bigr)
\biggr)
 $$ $$
 +{\lambda_1} {\tau} {\Theta_{23}} 
\biggl(
 {g^{33}} 
 \left( {R_2} {\Theta_{22}} {g^{23}}' - {R_3} {\Theta_{33}} {g^{22}}' \right)
$$ $$
 +{g^{22}} 
\Bigl(
 {\Theta_{33}} \left({R_3} {g^{33}}' - {R_2} {g^{23}}' \right)
 -2 {R_3} {g^{33}}^2
\Bigr)
$$ 
\begin{equation}
 +{R_2} {g^{23}} \left({\Theta_{33}} {g^{22}}' - 2 {g^{22}} {g^{33}} - {\Theta_{22}} {g^{33}}' \right)+2 {R_3} ({g^{23}})^2 {g^{33}} + 2 {R_2} {g^{23}}^3
\biggr)
\Biggr]
,\end{equation}
$$
\tilde V{}^3  
= 
\frac{1}{2 {\lambda_1} {\tau} \left({\Theta_{23}}^2-{\Theta_{22}} {\Theta_{33}}\right) \left(({g^{23}})^2-{g^{22}} {g^{33}} \right)}
\Biggl[
 {\lambda_1}^3 {\tau} {\vartheta_2} 
\biggl(
{\Theta_{23}} \left({g^{22}}' {g^{33}} - {g^{23}}  {g^{23}}' \right)
$$ $$
+{\Theta_{22}} \left({g^{23}} {g^{33}}' -{g^{23}}' {g^{33}} \right)
\biggr)
 $$ $$
 +{\lambda_1}^3 {\tau} {\vartheta_3} \left({\Theta_{22}} \left({g^{23}} {g^{23}}' - {g^{22}} {g^{33}}' \right)+{\Theta_{23}} \left({g^{22}} {g^{23}}' - {g^{22}}' {g^{23}}
  \right)\right)
 $$ $$
 +{\Theta_{22}} 
\biggl(
 {\Theta_{33}} \left({g^{22}} \left(2 {\lambda_3} {\rho_1} {g^{33}} - {\lambda_1} {\tau} {R_3} {g^{33}}' \right)
 -2 {\lambda_3} {\rho_1} ({g^{23}})^2+{\lambda_1} {\tau} {R_3} {g^{23}} {g^{23}}' \right)
$$ $$
 +{\lambda_1} {\tau} 
 \,
\Bigl(
 {g^{33}} \left(2 {R_3} {g^{22}} {g^{33}} - {R_2} {\Theta_{22}} {g^{23}}' \right)
 +{R_2} {g^{23}} \left(2 {g^{22}} {g^{33}} + {\Theta_{22}} {g^{33}}' \right)
 $$ $$
 -2 {R_3} ({g^{23}})^2 {g^{33}} - 2 {R_2} {g^{23}}^3
\Bigr)
\biggr)
 $$ $$
 +{\Theta_{23}}^2 
\Bigl(
 {g^{22}} \left({\lambda_1} {\tau} {R_2} {g^{23}}' - 2 {\lambda_3} {\rho_1} {g^{33}} \right)
 -{\lambda_1} {\tau} {g^{23}} \left({R_2} {g^{22}}' + {R_3} {g^{23}}' \right)
$$ $$
  +2 {\lambda_3} {\rho_1} ({g^{23}})^2
 +{\lambda_1} {\tau} {R_3} {g^{22}}' {g^{33}} 
\Bigr)
 $$ $$
 +{\lambda_1} {\tau} {\Theta_{23}} 
\biggl(
 {\Theta_{22}} {g^{33}} \left({R_2} {g^{22}}' - {R_3} {g^{23}}' \right)+{g^{23}} \left({R_3} {\Theta_{22}} {g^{33}}' - {R_3} {\Theta_{33}} {g^{22}}' \right)
 $$ $$
 +{g^{22}} \left(-{R_2} {\Theta_{22}} {g^{33}}' +{R_3} {\Theta_{33}} {g^{23}}' - 2 {R_3} {g^{23}} {g^{33}} + 2 {R_2} ({g^{23}})^2\right)
$$ 
\begin{equation}
 -2 {R_2} {g^{22}}^2 {g^{33}} + 2 {R_3} {g^{23}}^3
\biggr)
\Biggr]
,\end{equation}
%
%
%
where the functions $g^{ab}$ and $\Theta_{ab}$ (\ref{IntegralsOfMetric}) depend on one variable equal to the product of $\lambda_1\tau$, the variable $\tau$ is the time in the synchronous frame of reference, auxiliary constants $R_a$ are defined by the relations (\ref{R2R3}).

Due to the cumbersomeness of the expressions, the form of tidal acceleration 
$\tilde A{}^\alpha(\tau)=D^2\tilde \eta{}^\alpha/{d\tau}^2$
in the synchronous reference system is given
 in the appendix~(\ref{AppendixB}).
Note here that the tidal acceleration components $\tilde A{}^0$ and $\tilde A{}^1$ vanish in the used synchronous frame of reference:
\begin{equation}
\tilde A{}^0= 0 
,\qquad
\tilde A{}^1= 0 
, \end{equation}

We see that the tidal acceleration acts on particles only in the plane of the space variables $x^2$ and $x^3$, while the gravitational wave propagates along the space variable $x^1$.

Thus, for the considered models of gravitational waves, we obtained the trajectories of test particles, the geodesic deviation vector, the deviation velocity vector, and the form of tidal accelerations both in privileged and in synchronous coordinate systems. We have also found the structure of the gravitational wave metric in a synchronous frame of reference.

The characteristics of a gravitational wave obtained here completely determine its effect on particles, fields and the physical medium through which the wave passes. An example of the application of the obtained results to the calculation of secondary physical effects for a gravitational wave in the Bianchi Universe of type VII \cite{Osetrin2022894} is given to obtain 
retarded electromagnetic potentials of the Lienard-Wiechert type 
when charges move due to the tidal acceleration in a gravitational wave.

\section{Discussion}

%

The general results obtained in the paper on the deviation of geodesics in a gravitational wave can be used to calculate the physical effects of the influence of a gravitational wave on particles and fields. In particular, on the basis of the results obtained, one can calculate the radiation of charges moving with tidal acceleration in a gravitational wave and other secondary effects: the generation of density waves in a medium and the formation of local dense accumulations of matter, the birth of black holes in a gravitational wave \cite{Saito2010867, Saito200916}, the influence of primary gravitational waves on the microwave background of the universe and on the stochastic gravitational background, etc. 

Secondary effects allow one to estimate the parameters of the gravitational wave itself, especially for large wavelengths, when direct detection of the wave is impossible.
The gravitational waves under consideration contain special cases of spatially homogeneous nonisotropic Bianchi Universes of types IV, VI, and VII, on the basis of which, using the described approaches, exact models of primordial gravitational waves have already been 
constructed~\cite{Osetrin2022EPJP856,Osetrin2022894,https://doi.org/10.48550/arxiv.2209.08589} and the influence of the radiation generated by the primary plasma in the gravitational wave on the cosmic microwave background  parameters can be estimated \cite{Bennett2013}, including its observed anisotropy.

The exact solutions for gravitational waves obtained on the basis of the approaches proposed in this work can also be used to debug more complex computer models and train artificial intelligence systems to analyze gravitational wave signals.

\section{Conclusion}

%
%
%
%
%

For models of plane gravitational waves in Shapovalov spaces of type III, on the basis of the Hamilton-Jacobi formalism, exact solutions for the trajectories of test particles, the exact solution of geodesic deviation equations are obtained, the geodesic deviation velocities and tidal accelerations of a gravitational wave are found in a privileged coordinate system, where the spacetime metric is depends on the wave variable.

Using the obtained results, an explicit analytical transformation into a synchronous (laboratory) frame of reference associated with a freely falling observer on the base geodesic is found. The synchronous reference system allows you to synchronize time at different points in space and separate time and spatial variables, which is important for the observer and makes the resulting models physically illustrative.

In the synchronous reference system, the form of the metric, the geodesic deviation vector, the deviation velocity and tidal accelerations in a gravitational wave are found.

The obtained exact solutions make it possible to calculate the secondary physical effects by the action of gravitational waves.


\section*{Acknowledgments}

The authors thank the administration of the Tomsk State Pedagogical University for the technical support of the scientific project. 

The study was supported by the Russian Science Foundation, grant \mbox{No. 22-21-00265},
\url{https://rscf.ru/project/22-21-00265/}

\section*{Statements and Declarations}

\subsection*{Data availability statement}

All necessary data and references to external sources are contained in the text of the manuscript. 
All information sources used in the work are publicly available and refer to open publications in scientific journals and textbooks.

\subsection*{Compliance with Ethical Standards}

The authors declare no conflict of interest.

\subsection*{Competing Interests and Funding}
The study was supported by the Russian Science Foundation, 
\\grant \mbox{No. 22-21-00265}, 
\url{https://rscf.ru/project/22-21-00265/}








\appendix
\newpage
\section{
Tidal accelerations of a gravitational wave in a privileged coordinate system 
}
\label{AppendixA}

Below is a general form of tidal accelerations $A^\alpha(\tau)=D^2\eta^\alpha/{d\tau}^2$ acting on test particles in a gravitational wave with the metric $g^{\alpha\beta}(x^0)$ (\ref{PrivilMetricUp}) and the deviation vector $\eta^\alpha(\tau)$ (\ref{PrivilCommonEta0})-(\ref{PrivilCommonEta3}) in the privileged coordinate system in general case, without taking into account the field equations:
\begin{equation}
A^0 = 0
,\end{equation}
$$
A^1 (\tau) = 
\frac{
{\lambda_1} 
{\vartheta_3} 
 }{4 \left(({g^{23}})^2-{g^{22}} {g^{33}} \right)^2}
\Biggl[
2 \left({\lambda_2} {g^{22}}'' +{\lambda_3} {g^{23}}'' \right) {g^{23}}^3
$$ $$
-\left(3 {\lambda_3} {g^{23}}'{}^2+3 {g^{22}}' \left(2 {\lambda_2} {g^{23}}' +{\lambda_3} {g^{33}}' \right)+2 {g^{22}} \left({\lambda_2} {g^{23}}'' +{\lambda_3} {g^{33}}'' \right)\right) ({g^{23}})^2
$$ $$
+
{g^{23}} 
\biggl(
3 {g^{22}} \left(2 {\lambda_2} {g^{23}}'{}^2+3 {\lambda_3} {g^{33}}' {g^{23}}' +{\lambda_2} {g^{22}}' {g^{33}}' \right)
$$ $$
+{g^{33}} \left(3 {\lambda_2} {g^{22}}'{}^2+3 {\lambda_3} {g^{23}}' {g^{22}}' -2 {g^{22}} \left({\lambda_2} {g^{22}}'' +{\lambda_3} {g^{23}}'' \right)\right)
\biggr)
$$ $$
+{g^{22}} 
\biggl(
{g^{33}} \left(-3 {\lambda_3} {g^{23}}'{}^2-3 {\lambda_2} {g^{22}}' {g^{23}}' +2 {g^{22}} \left({\lambda_2} {g^{23}}'' +{\lambda_3} {g^{33}}'' \right)\right)
$$ $$
-3 {g^{22}} {g^{33}}' \left({\lambda_2} {g^{23}}' +{\lambda_3} {g^{33}}' \right)
\biggr)
\Biggr]
$$ $$
\mbox{}
-\frac{
{\lambda_1} 
{\vartheta_2} 
}{4 \left(({g^{23}})^2-{g^{22}} {g^{33}} \right)^2}
\Biggl[
\left(3 {\lambda_2} {g^{22}}'{}^2+3 {\lambda_3} {g^{23}}' {g^{22}}' -2 {g^{22}} \left({\lambda_2} {g^{22}}'' +{\lambda_3} {g^{23}}'' \right)\right) {g^{33}}^2
$$ $$
+
{g^{33}}
\biggl(
2 \left({\lambda_2} {g^{22}}'' +{\lambda_3} {g^{23}}'' \right) ({g^{23}})^2
+2 {g^{22}} \left({\lambda_2} {g^{23}}'' +{\lambda_3} {g^{33}}'' \right)
$$ $$
-\left( 6 {\lambda_3} {g^{23}}'{}^2+3 {g^{22}}' \left(3 {\lambda_2} {g^{23}}' 
+{\lambda_3} {g^{33}}' \right)
\right)
{g^{23}} 
+3 {g^{22}} {g^{23}}' \left({\lambda_2} {g^{23}}' +{\lambda_3} {g^{33}}' \right)
\biggr)
$$ $$
\mbox{}
 +{g^{23}} 
\biggl(
 -2 \left({\lambda_2} {g^{23}}'' +{\lambda_3} {g^{33}}'' \right) ({g^{23}})^2
  -3 {g^{22}} {g^{33}}' \left({\lambda_2} {g^{23}}' +{\lambda_3} {g^{33}}' \right)
 $$ $$
 \mbox{}
 +3 \left({\lambda_2} {g^{23}}'{}^2+2 {\lambda_3} {g^{33}}' {g^{23}}' +{\lambda_2} {g^{22}}' {g^{33}}' \right) {g^{23}} 
\biggr)
\Biggr]
$$ $$
\mbox{}
+
\frac{
1
}{4 {\lambda_1} \left(({g^{23}})^2-{g^{22}} {g^{33}} \right)^2}
\Biggl[
{R_3} {\Theta_{33}} 
\biggl(
2 \left(
{\lambda_2} {g^{22}}'' +{\lambda_3} {g^{23}}'' \right) ({g^{23}})^3
$$ $$
-\left(3 {\lambda_3} {g^{23}}'{}^2+3 {g^{22}}' \left(2 {\lambda_2} {g^{23}}' +{\lambda_3} {g^{33}}' \right)+2 {g^{22}} \left({\lambda_2} {g^{23}}'' +{\lambda_3} {g^{33}}'' \right)
\right) 
({g^{23}})^2
$$ $$
+
{g^{23}} 
\Bigl(
3 {g^{22}} \left(2 {\lambda_2} {g^{23}}'{}^2+3 {\lambda_3} {g^{33}}' {g^{23}}' +{\lambda_2} {g^{22}}' {g^{33}}' \right)
$$ $$
+{g^{33}} \left(3 {\lambda_2} {g^{22}}'{}^2+3 {\lambda_3} {g^{23}}' {g^{22}}' -2 {g^{22}} \left({\lambda_2} {g^{22}}'' +{\lambda_3} {g^{23}}'' \right)\right)
\Bigr)
$$ $$
+
{g^{22}} 
\Bigl(
{g^{33}} \left(-3 {\lambda_3} {g^{23}}'{}^2-3 {\lambda_2} {g^{22}}' {g^{23}}' +2 {g^{22}} \left({\lambda_2} {g^{23}}'' +{\lambda_3} {g^{33}}'' \right)\right)
$$ $$
-3 {g^{22}} {g^{33}}' \left({\lambda_2} {g^{23}}' +{\lambda_3} {g^{33}}' \right)
\Bigr)
\biggr)
$$ 
$$ 
-{R_2} {\Theta_{22}} 
\biggl(
\left(3 {\lambda_2} {g^{22}}'{}^2+3 {\lambda_3} {g^{23}}' {g^{22}}' -2 {g^{22}} \left({\lambda_2} {g^{22}}'' +{\lambda_3} {g^{23}}'' \right)\right) {g^{33}}^2
$$ 
$$
+
{g^{33}}
\Bigl(
2 \left({\lambda_2} {g^{22}}'' +{\lambda_3} {g^{23}}'' \right) ({g^{23}})^2
$$ $$
-\left(6 {\lambda_3} {g^{23}}'{}^2+3 {g^{22}}' \left(3 {\lambda_2} {g^{23}}' +{\lambda_3} {g^{33}}' \right)
-2 {g^{22}} \left({\lambda_2} {g^{23}}'' +{\lambda_3} {g^{33}}'' \right)
\right) 
{g^{23}} 
$$ $$
\mbox{}
+3 {g^{22}} {g^{23}}' \left({\lambda_2} {g^{23}}' +{\lambda_3} {g^{33}}' \right)
\Bigr)
$$ $$
+{g^{23}} 
\Bigl(
-2 \left({\lambda_2} {g^{23}}'' +{\lambda_3} {g^{33}}'' \right) ({g^{23}})^2
-3 {g^{22}} {g^{33}}' \left({\lambda_2} {g^{23}}' +{\lambda_3} {g^{33}}' \right)
$$ $$
+3 \left({\lambda_2} {g^{23}}'{}^2+2 {\lambda_3} {g^{33}}' {g^{23}}' +{\lambda_2} {g^{22}}' {g^{33}}' \right) {g^{23}} 
\Bigr)
\biggr)
$$ $$
+{\Theta_{23}} 
\biggl(
2 \left({\lambda_2} {R_2} {g^{22}}'' +({\lambda_3} {R_2}+{\lambda_2} {R_3}) {g^{23}}'' +{\lambda_3} {R_3} {g^{33}}'' \right) {g^{23}}^3
$$ 
$$
-
({g^{23}})^2
\Bigl[
3 ({\lambda_3} {R_2}+{\lambda_2} {R_3}) {g^{23}}'{}^2
+6 {\lambda_3} {R_3} {g^{33}}' {g^{23}}' 
+2 {R_2} {g^{22}} \left({\lambda_2} {g^{23}}'' +{\lambda_3} {g^{33}}'' \right)
$$
$$
+3 {g^{22}}' \left(2 {\lambda_2} {R_2} {g^{23}}' +({\lambda_3} {R_2}+{\lambda_2} {R_3}) {g^{33}}' \right)+2 {R_3} {g^{33}} \left({\lambda_2} {g^{22}}'' +{\lambda_3} {g^{23}}'' \right)
\Bigr]
$$ 
$$
\mbox{}
+
{g^{23}} 
\biggl(
3 {g^{22}} \left(2 {\lambda_2} {R_2} {g^{23}}'{}^2+(3 {\lambda_3} {R_2}+{\lambda_2} {R_3}) {g^{33}}' {g^{23}}' +{g^{33}}' \left({\lambda_2} {R_2} {g^{22}}' +{\lambda_3} {R_3} {g^{33}}' \right)\right)
$$ $$
+{g^{33}} 
\Bigl(
3 {\lambda_2} {R_2} {g^{22}}'{}^2+3 \left(({\lambda_3} {R_2}+3 {\lambda_2} {R_3}) {g^{23}}' +{\lambda_3} {R_3} {g^{33}}' \right) {g^{22}}' +6 {\lambda_3} {R_3} {g^{23}}'{}^2
$$ $$
-2 {g^{22}} \left({\lambda_2} {R_2} {g^{22}}'' +({\lambda_3} {R_2}+{\lambda_2} {R_3}) {g^{23}}'' +{\lambda_3} {R_3} {g^{33}}'' \right)
\Bigr)
\biggr)
$$ $$
-3 {R_2} {g^{22}}^2 {g^{33}}' \left({\lambda_2} {g^{23}}' +{\lambda_3} {g^{33}}' \right)
$$ $$
+{R_3} {g^{33}}^2 \left(-3 {\lambda_2} {g^{22}}'{}^2-3 {\lambda_3} {g^{23}}' {g^{22}}' +2 {g^{22}} \left({\lambda_2} {g^{22}}'' +{\lambda_3} {g^{23}}'' \right)\right)
$$ 
$$
+{g^{22}} {g^{33}} 
\Bigl(
-3 ({\lambda_3} {R_2}+{\lambda_2} {R_3}) {g^{23}}'{}^2
-3 {\lambda_2} {R_2} {g^{22}}' {g^{23}}' 
$$
\begin{equation}
-3 {\lambda_3} {R_3} {g^{33}}' {g^{23}}' 
+2 {R_2} {g^{22}} \left({\lambda_2} {g^{23}}'' +{\lambda_3} {g^{33}}'' \right)
\Bigr)
\biggr)
\Biggr]
,\end{equation} 

$$
A^2 (\tau) = 
 \frac{
  {\lambda_1}^2
 {\vartheta_3} 
  }{4 \left(({g^{23}})^2-{g^{22}} {g^{33}} \right)^2}
\Biggl[
 -2 {g^{22}}'' {g^{23}}^3+2 \left(3 {g^{22}}' {g^{23}}' +{g^{22}} {g^{23}}'' \right) ({g^{23}})^2
 $$ $$
 -\left(3 {g^{22}} \left(2 {g^{23}}'{}^2+{g^{22}}' {g^{33}}' \right)+{g^{33}} \left(3 {g^{22}}'{}^2-2 {g^{22}} {g^{22}}'' \right)\right) {g^{23}} 
 $$ $$
 +{g^{22}} \left(3 {g^{22}} {g^{23}}' {g^{33}}' +{g^{33}} \left(3 {g^{22}}' {g^{23}}' -2 {g^{22}} {g^{23}}'' \right)\right)
\Biggr]
$$
$$
\mbox{}
+
\frac{
{\lambda_1}^2
{\vartheta_2} 
}{4 \left(({g^{23}})^2-{g^{22}} {g^{33}} \right)^2}
\Biggl[
\left(3 {g^{22}}'{}^2-2 {g^{22}} {g^{22}}'' \right) {g^{33}}^2
$$ $$
+\left(
2 {g^{22}}'' ({g^{23}})^2+\left(2 {g^{22}} {g^{23}}'' -9 {g^{22}}' {g^{23}}' \right) {g^{23}} 
+3 {g^{22}} {g^{23}}'{}^2
\right) 
{g^{33}} 
$$ $$
+{g^{23}} \left(-2 {g^{23}}'' ({g^{23}})^2+3 \left({g^{23}}'{}^2+{g^{22}}' {g^{33}}' \right) {g^{23}} -3 {g^{22}} {g^{23}}' {g^{33}}' \right)
\Biggr]
$$ $$
+
\frac{
1
}{4 \left(({g^{23}})^2-{g^{22}} {g^{33}} \right)^2}
\Biggl[
{R_3} {\Theta_{33}} 
\biggl(
-2 {g^{22}}'' {g^{23}}^3+2 \left(3 {g^{22}}' {g^{23}}' +{g^{22}} {g^{23}}'' \right) ({g^{23}})^2
$$ $$
-\left(3 {g^{22}} \left(2 {g^{23}}'{}^2+{g^{22}}' {g^{33}}' \right)+{g^{33}} \left(3 {g^{22}}'{}^2-2 {g^{22}} {g^{22}}'' \right)\right) {g^{23}} 
$$ $$
+{g^{22}} \left(3 {g^{22}} {g^{23}}' {g^{33}}' +{g^{33}} \left(3 {g^{22}}' {g^{23}}' -2 {g^{22}} {g^{23}}'' \right)\right)
\biggr)
$$ $$
+
{R_2} {\Theta_{22}} 
\biggl(
\left(3 {g^{22}}'{}^2-2 {g^{22}} {g^{22}}'' \right) {g^{33}}^2
$$ $$
+\left(2 {g^{22}}'' ({g^{23}})^2+\left(2 {g^{22}} {g^{23}}'' -9 {g^{22}}' {g^{23}}' \right) {g^{23}} +3 {g^{22}} {g^{23}}'{}^2\right) {g^{33}} 
$$ $$
+{g^{23}} \left(-2 {g^{23}}'' ({g^{23}})^2+3 \left({g^{23}}'{}^2+{g^{22}}' {g^{33}}' \right) {g^{23}} -3 {g^{22}} {g^{23}}' {g^{33}}' \right)
\biggr)
$$ $$
+{\Theta_{23}} 
\biggl(
-2 \left({R_2} {g^{22}}'' +{R_3} {g^{23}}'' \right) {g^{23}}^3
$$ $$
+\left(3 {R_3} {g^{23}}'{}^2+3 {g^{22}}' \left(2 {R_2} {g^{23}}' +{R_3} {g^{33}}' \right)+2 {R_3} {g^{33}} {g^{22}}'' +2 {R_2} {g^{22}} {g^{23}}'' \right) ({g^{23}})^2
$$ $$
\mbox{}
-
{g^{23}} 
\Bigl[
3 {g^{22}} \left(2 {R_2} {g^{23}}'{}^2+{R_3} {g^{33}}' {g^{23}}' +{R_2} {g^{22}}' {g^{33}}' \right)
$$ $$
+{g^{33}} \left(3 {R_2} {g^{22}}'{}^2+9 {R_3} {g^{23}}' {g^{22}}' -2 {g^{22}} \left({R_2} {g^{22}}'' +{R_3} {g^{23}}'' \right)\right)
\Bigr]
$$
$$
+3 {R_2} {g^{22}}^2 {g^{23}}' {g^{33}}' +{R_3} {g^{33}}^2 \left(3 {g^{22}}'{}^2-2 {g^{22}} {g^{22}}'' \right)
$$ 
\begin{equation}
\mbox{}
+{g^{22}} {g^{33}} \left(3 {R_3} {g^{23}}'{}^2+3 {R_2} {g^{22}}' {g^{23}}' -2 {R_2} {g^{22}} {g^{23}}'' \right)
\biggr)
\Biggr]
,\end{equation}

$$
A^3 (\tau)
 = \frac{
  {\lambda_1}^2
 {\vartheta_3} 
 }{4 \left(({g^{23}})^2-{g^{22}} {g^{33}} \right)^2}
\Biggl[
 -2 {g^{23}}'' {g^{23}}^3+\left(3 {g^{23}}'{}^2+3 {g^{22}}' {g^{33}}' +2 {g^{22}} {g^{33}}'' \right) ({g^{23}})^2
 $$ $$
 +\left({g^{33}} \left(2 {g^{22}} {g^{23}}'' -3 {g^{22}}' {g^{23}}' \right)-9 {g^{22}} {g^{23}}' {g^{33}}' \right) {g^{23}} 
 $$ $$
 +{g^{22}} \left(3 {g^{22}} {g^{33}}'{}^2+{g^{33}} \left(3 {g^{23}}'{}^2-2 {g^{22}} {g^{33}}'' \right)\right)
\Biggr]
$$ $$
\mbox{}
+\frac{
{\lambda_1}^2
{\vartheta_2} 
}{4 \left(({g^{23}})^2-{g^{22}} {g^{33}} \right)^2}
\Biggl[
\left(3 {g^{22}}' {g^{23}}' -2 {g^{22}} {g^{23}}'' \right) {g^{33}}^2
$$ $$
+\left(2 {g^{23}}'' ({g^{23}})^2+\left(-6 {g^{23}}'{}^2-3 {g^{22}}' {g^{33}}' +2 {g^{22}} {g^{33}}'' \right) {g^{23}} +3 {g^{22}} {g^{23}}' {g^{33}}' \right) {g^{33}} 
$$ $$
+{g^{23}} \left(-2 {g^{33}}'' ({g^{23}})^2+6 {g^{23}}' {g^{33}}' {g^{23}} -3 {g^{22}} {g^{33}}'{}^2\right)
\Biggr]
$$ 
$$
\mbox{}
+
\frac{
1
}{4 \left(({g^{23}})^2-{g^{22}} {g^{33}} \right)^2}
\Biggl[
{R_3} {\Theta_{33}} 
\biggl(
-2 {g^{23}}'' {g^{23}}^3
+\left(3 {g^{23}}'{}^2+3 {g^{22}}' {g^{33}}' +2 {g^{22}} {g^{33}}'' \right) ({g^{23}})^2
$$ $$
+\left({g^{33}} \left(2 {g^{22}} {g^{23}}'' -3 {g^{22}}' {g^{23}}' \right)-9 {g^{22}} {g^{23}}' {g^{33}}' \right) {g^{23}} 
$$ $$
+{g^{22}} \left(3 {g^{22}} {g^{33}}'{}^2+{g^{33}} \left(3 {g^{23}}'{}^2-2 {g^{22}} {g^{33}}'' \right)\right)
\biggr)
$$ $$
\mbox{}
+
{R_2} {\Theta_{22}} 
\biggl(
\left(3 {g^{22}}' {g^{23}}' -2 {g^{22}} {g^{23}}'' \right) {g^{33}}^2
$$
$$
+\left(2 {g^{23}}'' ({g^{23}})^2+\left(-6 {g^{23}}'{}^2-3 {g^{22}}' {g^{33}}' +2 {g^{22}} {g^{33}}'' \right) {g^{23}} +3 {g^{22}} {g^{23}}' {g^{33}}' \right) {g^{33}} 
$$
$$
+{g^{23}} \left(-2 {g^{33}}'' ({g^{23}})^2+6 {g^{23}}' {g^{33}}' {g^{23}} -3 {g^{22}} {g^{33}}'{}^2\right)
\biggr)
$$ $$
+
{\Theta_{23}} 
\biggl(
-2 {g^{23}}^3 \left({R_2} {g^{23}}'' +{R_3} {g^{33}}'' \right) 
$$ $$
+\left(3 {R_2} {g^{23}}'{}^2+6 {R_3} {g^{33}}' {g^{23}}' 
+3 {R_2} {g^{22}}' {g^{33}}' +2 {R_3} {g^{33}} {g^{23}}'' +2 {R_2} {g^{22}} {g^{33}}'' \right) ({g^{23}})^2
$$ $$
-
\Bigl(
3 {g^{22}} {g^{33}}' \left(3 {R_2} {g^{23}}' +{R_3} {g^{33}}' \right)
$$ $$
+{g^{33}} \left(6 {R_3} {g^{23}}'{}^2+3 {g^{22}}' \left({R_2} {g^{23}}' +{R_3} {g^{33}}' \right)-2 {g^{22}} \left({R_2} {g^{23}}'' +{R_3} {g^{33}}'' \right)\right)
\Bigr) {g^{23}} 
$$ $$
+3 {R_2} {g^{22}}^2 {g^{33}}'{}^2+{R_3} {g^{33}}^2 \left(3 {g^{22}}' {g^{23}}' -2 {g^{22}} {g^{23}}'' \right)
$$
\begin{equation}
+{g^{22}} {g^{33}} \left(3 {R_2} {g^{23}}'{}^2+3 {R_3} {g^{33}}' {g^{23}}' -2 {R_2} {g^{22}} {g^{33}}'' \right)
\biggr)
\Biggr]
,\end{equation} 

\newpage

\section{
Tidal accelerations of a gravitational wave in a synchronous frame of reference (taking into account Einstein's vacuum equations)
}
\label{AppendixB}

Below is the form of tidal accelerations $\tilde A{}^\alpha(\tau)=D^2\tilde\eta{}^\alpha/{d\tau}^2$ acting on test particles in a gravitational wave
with metric $\tilde g{}^{\alpha\beta}(\tau,\tilde x{}^k)$ (\ref{SynchrMetricUp0k})-(\ref{SynchrMetricUpab})
and deviation vector $\tilde \eta{}^\alpha(\tau)$
taking into account the vacuum Einstein equation (\ref{EqR00})
in the synchronous reference system:
\begin{equation}
\tilde A{}^0= 0 
,\qquad
\tilde A{}^1= 0 
, \end{equation}
$$
\tilde A{}^2 (\tau) =
\frac{
{\lambda_1}^3
{\vartheta_3} 
}{4 \left({\Theta_{23}}{}^2-{\Theta_{22}} {\Theta_{33}} \right) \left(({g^{23}})^2-{g^{22}} {g^{33}} \right)^2}
\,
\Biggl[
{\Theta_{33}} 
\biggl(
2 {g^{22}}'' ({g^{23}})^3
$$
$$
+\left(3 {g^{22}} \left(2 {g^{23}}'{}^2+{g^{22}}' {g^{33}}' \right)
+{g^{33}} \left(3
   {g^{22}}'{}^2-2 {g^{22}} {g^{22}}'' \right)\right) {g^{23}}
$$
$$
+{g^{22}} \left({g^{33}} \left(2 {g^{22}}{}   {g^{23}}'' -3 {g^{22}}' {g^{23}}' \right)-3 {g^{22}} {g^{23}}' {g^{33}}' \right)
$$ $$
-2 \left(3 {g^{22}}' {g^{23}}' +{g^{22}} {g^{23}}'' \right) ({g^{23}})^2
\biggr)
$$ $$
+
{\Theta_{23}}
\biggl(
-2 {g^{23}}'' ({g^{23}})^3+\left(3 {g^{23}}'{}^2+3 {g^{22}}' {g^{33}}' +2 {g^{22}} {g^{33}}'' \right)
   ({g^{23}})^2
$$ $$
+\left({g^{33}} \left(2 {g^{22}} {g^{23}}'' -3 {g^{22}}' {g^{23}}' \right)-9 {g^{22}}{}
   {g^{23}}' {g^{33}}' \right) {g^{23}} 
$$ $$
+{g^{22}}{}   \left(3 {g^{22}} {g^{33}}'{}^2+{g^{33}} \left(3 {g^{23}}'{}^2-2 {g^{22}} {g^{33}}'' \right)\right)
\,
\biggr)
\Biggr]
$$ 
$$
-\frac{
{\lambda_1}^3
{\vartheta_2} 
}{4 \left({\Theta_{23}}{}^2-{\Theta_{22}} {\Theta_{33}} \right) \left(({g^{23}})^2-{g^{22}}{g^{33}} \right)^2}
\Biggl[
{\Theta_{33}} 
\biggl(
\left(3 {g^{22}}'{}^2-2 {g^{22}} {g^{22}}'' \right) ({g^{33}})^2
$$ 
$$
+\left(2 {g^{22}}'' ({g^{23}})^2+\left(2 {g^{22}} {g^{23}}'' -9 {g^{22}}' {g^{23}}' \right)
   {g^{23}} +3 {g^{22}} {g^{23}}'{}^2\right) {g^{33}} 
$$ 
$$
+{g^{23}} \left(-2 {g^{23}}'' ({g^{23}})^2+3
   \left({g^{23}}'{}^2+{g^{22}}' {g^{33}}' \right) {g^{23}} -3 {g^{22}} {g^{23}}' {g^{33}}' \right)
\biggr)
$$ 
$$
+
{\Theta_{23}} 
\biggl(
\left(2 {g^{22}} {g^{23}}'' -3  {g^{22}}' {g^{23}}' \right) ({g^{33}})^2
$$ 
$$
+
{g^{33}}
\Bigl(
{g^{23}}
\left(6 {g^{23}}'{}^2+3 {g^{22}}'{}   {g^{33}}' -2 {g^{22}} {g^{33}}'' \right)  
$$ $$
-2 {g^{23}}'' ({g^{23}})^2
-3 {g^{22}} {g^{23}}' {g^{33}}' 
\Bigr)    
$$ 
$$
+{g^{23}} \left(2 {g^{33}}'' ({g^{23}})^2-6 {g^{23}}' {g^{33}}' {g^{23}} +3 {g^{22}} {g^{33}}'{}^2\right)
\biggr)
\Biggr] 
$$ $$
+
\frac{
{\lambda_1} 
{R_3} 
}{4 \left({\Theta_{23}}{}^2-{\Theta_{22}}   {\Theta_{33}} \right) \left(({g^{23}})^2-{g^{22}} {g^{33}} \right)^2}
\Biggl[
{\Theta_{23}}{}^2
\Biggl(
\left(3 {g^{22}}' {g^{23}}' -2   {g^{22}} {g^{23}}'' \right) ({g^{33}})^2
$$ $$
+\left(2 {g^{23}}'' ({g^{23}})^2+\left(-6 {g^{23}}'{}^2-3 {g^{22}}'{}   {g^{33}}' +2 {g^{22}} {g^{33}}'' \right) {g^{23}} +3 {g^{22}} {g^{23}}' {g^{33}}' \right)   {g^{33}} 
$$ $$
+{g^{23}} \left(-2 {g^{33}}'' ({g^{23}})^2+6 {g^{23}}' {g^{33}}' {g^{23}} -3 {g^{22}} {g^{33}}'{}^2\right)
\Biggr) 
$$ $$
+
{\Theta_{23}} 
{\Theta_{33}} 
\left(
2   {g^{22}} {g^{22}}'' -3 {g^{22}}' {g^{23}} +3   {g^{22}} {g^{33}}'{}^2
\right) 
$$ $$
+{\Theta_{33}}{}^2 
\biggl(
2 {g^{22}}'' ({g^{23}})^3-2 \left(3 {g^{22}}'   {g^{23}}' +{g^{22}} {g^{23}}'' \right) ({g^{23}})^2
$$ $$
+\left(3 {g^{22}} \left(2 {g^{23}}'{}^2
+{g^{22}}'{}   {g^{33}}' \right)
+{g^{33}} \left(3 {g^{22}}'{}^2-2 {g^{22}} {g^{22}}'' \right)\right) {g^{23}} 
$$ $$
+{g^{22}}   \left({g^{33}} \left(2 {g^{22}} {g^{23}}'' -3 {g^{22}}' {g^{23}}' \right)
-3 {g^{22}} {g^{23}}'{}   {g^{33}}' 
\right)
\biggr)
\Biggr]
$$ $$
+
\frac{
{\lambda_1}
{R_2} 
}{4 \left({\Theta_{23}}{}^2-{\Theta_{22}} {\Theta_{33}}\right) \left(({g^{23}})^2-{g^{22}} {g^{33}} \right)^2}
\Biggl[
{\Theta_{23}}{}^2
\biggl(
-2 {g^{23}}'' ({g^{23}})^3
$$ $$
+\left(3 {g^{23}}'{}^2+3 {g^{22}}' {g^{33}}' +2 {g^{22}} {g^{33}}'' \right) ({g^{23}})^2
$$ $$
+\left({g^{33}} \left(2 {g^{22}} {g^{23}}'' -3 {g^{22}}' {g^{23}}' \right)
- 9   {g^{22}} {g^{23}}' {g^{33}}' \right) {g^{23}} 
$$ $$
+{g^{22}} \left(3 {g^{22}} {g^{33}}'{}^2+{g^{33}} \left(3 {g^{23}}'{}^2-2 {g^{22}}{}
   {g^{33}}'' 
   \right)\right)
\biggr)
$$ $$
+
{\Theta_{23}} 
\Biggl(
{\Theta_{33}} 
\biggl(
2 {g^{22}}'' ({g^{23}})^3
-2 
\left(
3 {g^{22}}' {g^{23}}' +{g^{22}} {g^{23}}'' \right) ({g^{23}})^2
$$ $$
+
\left(
3   {g^{22}} \left(2 {g^{23}}'{}^2+{g^{22}}' {g^{33}}' \right)
+{g^{33}} \left(3 {g^{22}}'{}^2-2 {g^{22}}{}   {g^{22}}'' \right)
\right) {g^{23}} 
$$ $$
+{g^{22}} \left({g^{33}} \left(2 {g^{22}} {g^{23}}'' -3 {g^{22}}'{}   {g^{23}}' \right)
-3 {g^{22}} {g^{23}}' {g^{33}}' \right)
\biggr)
$$ $$
+{\Theta_{22}} 
\biggl(
\left(3 {g^{22}}' {g^{23}}' -2   {g^{22}} {g^{23}}'' \right) ({g^{33}})^2
$$ $$
+
{g^{33}}
\left(
2 {g^{23}}'' ({g^{23}})^2
+\left(
3 {g^{22}} {g^{23}}' {g^{33}}' 
-6 {g^{23}}'{}^2-3 {g^{22}}' {g^{33}}' 
+2 {g^{22}} {g^{33}}'' \right) {g^{23}} 
\right)   
$$ $$
+{g^{23}} \left(-2 {g^{33}}'' ({g^{23}})^2+6 {g^{23}}' {g^{33}}' {g^{23}} -3 {g^{22}} {g^{33}}'^2\right)
\biggr)
\Biggr)
$$ $$
 -{\Theta_{22}}  {\Theta_{33}} 
\biggl(
 \left(3 {g^{22}}'{}^2-2 {g^{22}} {g^{22}}'' \right) ({g^{33}})^2
$$ $$
+\left(2 {g^{22}}'' ({g^{23}})^2
 +\left(2   {g^{22}} {g^{23}}'' -9 {g^{22}}' {g^{23}}' \right) {g^{23}} 
 +3 {g^{22}} {g^{23}}'{}^2\right)   {g^{33}} 
$$ 
\begin{equation}
+{g^{23}} \left(-2 {g^{23}}'' ({g^{23}})^2+3 \left({g^{23}}'{}^2+{g^{22}}' {g^{33}}' \right)
   {g^{23}} -3 {g^{22}} {g^{23}}' {g^{33}}'
   \right)
 \,
\biggr)
\Biggr] 
,\end{equation}
$$
\tilde A{}^3 (\tau) =
-\frac{
{\lambda_1}^3
{\vartheta_3} 
}{4 \left({\Theta_{23}}{}^2-{\Theta_{22}} {\Theta_{33}} \right) \left(({g^{23}})^2-{g^{22}} {g^{33}} \right)^2}
\Biggl[
{\Theta_{23}} 
\biggl(
2 {g^{22}}'' ({g^{23}})^3
$$ $$
-2 \left(3   {g^{22}}' {g^{23}}' +{g^{22}} {g^{23}}'' \right) ({g^{23}})^2
$$ $$
+\left(3 {g^{22}} \left(2 {g^{23}}'{}^2+{g^{22}}' {g^{33}}' \right)+{g^{33}} \left(3
   {g^{22}}'{}^2-2 {g^{22}} {g^{22}}'' \right)\right) {g^{23}} 
$$ $$
   +{g^{22}} \left({g^{33}} \left(2 {g^{22}}{}
   {g^{23}}'' -3 {g^{22}}' {g^{23}}' \right)-3 {g^{22}} {g^{23}}' {g^{33}}' \right)
\biggr)
$$ $$
+{\Theta_{22}} 
\biggl(
-2 {g^{23}}'' ({g^{23}})^3+\left(3 {g^{23}}'{}^2+3 {g^{22}}' {g^{33}}' +2 {g^{22}} {g^{33}}'' \right)
   ({g^{23}})^2
$$ $$
+\left({g^{33}} \left(2 {g^{22}} {g^{23}}'' -3 {g^{22}}' {g^{23}}' \right)-9 {g^{22}}{}
   {g^{23}}' {g^{33}}' \right) {g^{23}}
$$ $$
+{g^{22}}   \left(3 {g^{22}} {g^{33}}'{}^2
+{g^{33}} \left(3 {g^{23}}'{}^2-2 {g^{22}} {g^{33}}'' \right)\right)
\biggr)
\Biggr]
$$ $$
+
\frac{
{\lambda_1}^3
{\vartheta_2} 
}{4 \left({\Theta_{23}}{}^2-{\Theta_{22}} {\Theta_{33}} \right) \left(({g^{23}})^2-{g^{22}} {g^{33}} \right)^2}
\Biggl[
{\Theta_{23}} 
\biggl(
\left(3 {g^{22}}'{}^2-2 {g^{22}} {g^{22}}'' \right)   ({g^{33}})^2
$$  $$
+\left(2 {g^{22}}'' ({g^{23}})^2+\left(2 {g^{22}} {g^{23}}'' -9 {g^{22}}' {g^{23}}' \right)
   {g^{23}} +3 {g^{22}} {g^{23}}'{}^2\right) {g^{33}} 
$$ $$
+{g^{23}} \left(-2 {g^{23}}'' ({g^{23}})^2+3
   \left({g^{23}}'{}^2+{g^{22}}' {g^{33}}' \right) {g^{23}} -3 {g^{22}} {g^{23}}' {g^{33}}' \right)
\biggr)
$$ $$
+
{\Theta_{22}} 
\biggl(
\left(2 {g^{22}} {g^{23}}'' -3
   {g^{22}}' {g^{23}}' \right) ({g^{33}})^2
$$ $$
+\left(-2 {g^{23}}'' ({g^{23}})^2+\left(6 {g^{23}}'{}^2+3 {g^{22}}'{}   {g^{33}}' -2 {g^{22}} {g^{33}}'' \right) {g^{23}} -3 {g^{22}} {g^{23}}' {g^{33}}' \right)   {g^{33}}
$$ $$
 +{g^{23}} \left(2 {g^{33}}'' ({g^{23}})^2-6 {g^{23}}' {g^{33}}' {g^{23}} +3 {g^{22}} {g^{33}}'{}^2\right)
 \,
\biggr)
\Biggr] 
$$ 
$$
+\frac{
{\lambda_1}
{R_2} 
}{4 \left({\Theta_{23}}{}^2-{\Theta_{22}} {\Theta_{33}} \right) \left(({g^{23}})^2-{g^{22}} {g^{33}} \right)^2}
\Biggl[
{\Theta_{22}}{}^2
\biggl(
\left(2 {g^{22}} {g^{23}}'' -3   {g^{22}}' {g^{23}}' \right) ({g^{33}})^2
$$ $$
+\left(-2 {g^{23}}'' ({g^{23}})^2+\left(6 {g^{23}}'{}^2+3 {g^{22}}'{}   {g^{33}}' -2 {g^{22}} {g^{33}}'' \right) {g^{23}} -3 {g^{22}} {g^{23}}' {g^{33}}' \right)   {g^{33}} 
$$ $$
+{g^{23}} \left(2 {g^{33}}'' ({g^{23}})^2-6 {g^{23}}' {g^{33}}' {g^{23}} +3 {g^{22}} {g^{33}}'{}^2\right)
\biggr) 
$$ $$
+
{\Theta_{22}} 
{\Theta_{23}} 
\biggl(
\left(3 {g^{22}}'{}^2-2 {g^{22}} {g^{22}}'' \right) ({g^{33}})^2
$$ $$
+2 \left({g^{33}}'' ({g^{22}})^2-3 {g^{23}}   {g^{22}}' {g^{23}}' 
+({g^{23}})^2 {g^{22}}'' \right) {g^{33}} 
$$ $$
+{g^{22}} \left(-2 {g^{33}}''{}
   ({g^{23}})^2+6 {g^{23}}' {g^{33}}' {g^{23}} -3
   {g^{22}} {g^{33}}'{}^2\right)
\biggr)
$$ $$
+{\Theta_{23}}{}^2 
\biggl(
-2 {g^{22}}'' ({g^{23}})^3
+2 \left(3 {g^{22}}'{}   {g^{23}}' +{g^{22}} {g^{23}}'' \right) ({g^{23}})^2
$$ $$
-
{g^{23}} 
\left(
3 {g^{22}} \left(2 {g^{23}}'{}^2+{g^{22}}'{}  {g^{33}}' \right)
+{g^{33}} \left(3 {g^{22}}'{}^2-2 {g^{22}} {g^{22}}'' \right)
\right) 
$$ $$
+
{g^{22}} 
\left(
3   {g^{22}} {g^{23}}' {g^{33}}'
+{g^{33}} \left(3{g^{22}}' {g^{23}}' -2 {g^{22}} {g^{23}}'' \right)
\right)
\biggr)
\Biggr]
$$ 
$$
+
\frac{
{\lambda_1}
{R_3} 
}{4 \left({\Theta_{23}}{}^2-{\Theta_{22}} {\Theta_{33}} \right) \left( ({g^{23}})^2-{g^{22}} {g^{33}} \right)^2}
\Biggl[
{\Theta_{23}}{}^2
\biggl(
\left(3 {g^{22}}'{}^2-2 {g^{22}} {g^{22}}'' \right) ({g^{33}})^2
$$ $$
+\left(2 {g^{22}}'' ({g^{23}})^2+\left(2
   {g^{22}} {g^{23}}'' -9 {g^{22}}' {g^{23}}' \right) {g^{23}} +3 {g^{22}} {g^{23}}'{}^2\right)
   {g^{33}} 
$$ $$
   +{g^{23}} \left(-2 {g^{23}}'' ({g^{23}})^2+3 \left({g^{23}}'{}^2+{g^{22}}' {g^{33}}' \right)
   {g^{23}} -3 {g^{22}} {g^{23}}' {g^{33}}' \right)
\biggr)
$$ 
$$
+
{\Theta_{23}} 
{\Theta_{33}} 
\biggl(
-2 {g^{22}}'' ({g^{23}})^3+2 \left(3 {g^{22}}' {g^{23}}' +{g^{22}} {g^{23}}'' \right) ({g^{23}})^2
$$ $$
-\left(3 {g^{22}} \left(2   {g^{23}}'{}^2+{g^{22}}' {g^{33}}' \right)
   +{g^{33}} \left(3 {g^{22}}'{}^2-2 {g^{22}} {g^{22}}'' \right)\right)  {g^{23}} 
$$ $$
+{g^{22}} \left(3 {g^{22}} {g^{23}}' {g^{33}}' 
+{g^{33}} \left(3 {g^{22}}' {g^{23}}' -2 {g^{22}} {g^{23}}'' \right)\right)
\,
\biggr)
$$ $$
+
{\Theta_{22}}
{\Theta_{23}} 
\biggl(
\left(2 {g^{22}} {g^{23}}'' -3 {g^{22}}' {g^{23}}' \right) ({g^{33}})^2
$$ $$
-\left(2 {g^{23}}'' ({g^{23}})^2
-\left(6 {g^{23}}'{}^2+3 {g^{22}}' {g^{33}}' -2 {g^{22}}{}
   {g^{33}}'' \right) {g^{23}} 
+3 {g^{22}} {g^{23}}' {g^{33}}' \right) {g^{33}} 
$$ $$
+{g^{23}} \left(2  {g^{33}}'' ({g^{23}})^2-6 {g^{23}}' {g^{33}}' {g^{23}} +3 {g^{22}} {g^{33}}'{}^2\right)
\biggr)
$$ $$
+{\Theta_{22}} {\Theta_{33}} 
\biggl(
2 {g^{23}}''   ({g^{23}})^3-\left(3 {g^{23}}'{}^2+3 {g^{22}}' {g^{33}}' +2 {g^{22}} {g^{33}}'' \right) ({g^{23}})^2
$$
$$
+\left(9   {g^{22}} {g^{23}}' {g^{33}}' +{g^{33}} \left(3
   {g^{22}}' {g^{23}}' -2 {g^{22}} {g^{23}}'' \right)\right) {g^{23}} 
$$ 
\begin{equation} 
+{g^{22}} \left({g^{33}} \left(2   {g^{22}} {g^{33}}'' -3 {g^{23}}'{}^2\right)-3 {g^{22}} {g^{33}}'{}^2
\biggr)
\right)
\Biggl]
,\end{equation}
here the one-variable functions $g^{ab}$ and $\Theta_{ab}$ depend on the product $\lambda_1\tau$, where $\tau$ is the time in the synchronous reference frame.

%
%
%
%
%
%
%
%


\end{document}